# Canine Olfactory Differentiation of Cancer – A Review of the Literature

Oliver Gould[1], Amy Smart[1], Norman Ratcliffe[1], Ben de Lacy Costello[1],

[1]*Institute of Biosensor Technology, University of the West of England, Bristol, BS 16 1QY, United Kingdom*

Abstract

Numerous studies have attempted to demonstrate the olfactory ability of canines to detect several common cancer types from human bodily fluids, breath and tissue. Canines have been reported to detect bladder cancer (sensitivity of 0.63-0.73 and specificity of 0.64-0.92) and prostate cancer (sensitivity of 0.91-0.99 and specificity of 0.91-0.97) from urine; breast cancer (sensitivity of 0.88 and specificity of 0.98) and lung cancer (sensitivity 0.56-0.99 and specificity of 8.30-0.99) on breath and colorectal cancer from stools (sensitivity of 0.91-0.97 and specificity of 0.97-0.99). The quoted figures of sensitivity and specificity across differing studies demonstrate that in many cases results are variable from study to study; this raises questions about the reproducibility of methodology and study design which we have identified herein. Furthermore in some studies the controls used have resulted in differentiation of samples which are of limited use for clinical diagnosis.

These studies provide some evidence that cancer gives rise to different volatile organic compounds (VOCs) compared to healthy samples. Whilst canine detection may be unsuitable for clinical implementation they can, at least, provide inspiration for more traditional laboratory investigations.

Introduction

The olfactory ability of canines has long been recognised, and this ability has been exploited by humans for centuries e.g. bloodhounds for tracking people and animals. Since the on-set of World War II canines have been utilized to detect explosives, due in no small part to their rapid detection capabilities (Furton and Myers, 2001). Humans are estimated to have between 650 and 900 olfactory receptor genes, the canine genome contains an estimated 1300 (Quignon *et al*, 2005). While this difference does not at first inspection appear highly significant the human receptor gene repertoire is thought to contain a significantly higher proportion of pseudogenes compared to that of canines. As a result the human nose is thought to contain approximately 5 million olfactory receptors, whilst a canine nose is thought to contain in excess of forty times this figure (Lippi and Cervellin, 2012).

The use of canines for disease detection is a relatively new phenomenon. It is often cited that the first report of a canine's ability to 'detect' cancer was a letter to the Lancet from William and Pembroke (1989) which details the anecdotal account of a woman whose pet obsessed over a skin lesion, prompting her to seek medical attention. The lesion was later found to be a melanoma. Since then research into the area of disease detection by canines has rapidly expanded and scientists have attempted to rigorously test the olfactory ability of canines and other animals to detect odours associated with cancer. An early review by Moser and McCulloch (2010) provides an in-depth description and discussion of such studies, but is now outdated due to the rapidly expanding nature of this field. A later review by Bijland *et al*. (2013) broadly describes olfactory detection of disease by animals, although does not describe the field of canine detection comprehensively.

Recently, studies using canines to detect the infectious pathogenic *Clostridium difficile* in hospitals have received publicity. Canines have also been reported to have detected the gastrointestinal bacterial infection from stool samples with very high sensitivity and specificity (Bomers *et al.* 2012 and 2014), and have the advantage of being much quicker than current conventional methods (cytotoxin assay results take 1-2 days). This result is currently superior to that achieved by a prototype gas chromatography-gas sensor system (McGuire *et al,* 2014). Medical detection dogs have also been trained to sense oncoming hypoglycaemic episodes – a dangerous complication affecting those living with diabetes. It has been hypothesised canines can react to olfactory cues from changes in blood sugar levels; as the canines were able to alert their owners to hypo/hyper-glycaemic episodes even during sleep (Rooney *et al*, 2013).

In the western world, cancer is a growing problem probably due to an ageing population, which is not only of great cost to individuals, but puts a financial strain on healthcare services. Cancer prognosis and survival rates largely depend on the stage at which it is detected, with early diagnosis allowing more treatment options and better outcomes (Cancer Research UK, 2014). Where routine screening is available e.g. for breast (mammograms) and cervical cancer (smear tests) prognosis has drastically improved; however there are insufficient screening options available in most cases. To be effective, diagnostic tests must meet certain criterion including a need for high sensitivity and specificity. Additionally tests of this nature must be cost effective (cost of testing to incidence of disease ratio), provide high positive predictive values (PPV) and negative predictive values (NPV), ideally yield a rapid result, and biomarkers should correlate with disease progression and regression.

This field inspires many applications, namely the non-invasive screening of cancer using breath, urine, faeces etc. Whilst canines in a clinical setting may be impractical, by identifying the important compounds that canines can detect, there is the potential to optimise sampling methods and analytical techniques of these compounds with a view to developing the potential use of non-invasive biomarkers. There is also potential in the shorter term that canines could provide a service in second line screening where samples are brought to a canine in a separate location similar to the government licensed DSTL RASCargO™ system for explosive detection (Wickens, 2001). It should be highlighted though, that whilst some studies support the use of canines in detecting odours associated with cancers there are often large discrepancies between studies. For instance Ehmann *et al* (2012) investigating lung cancer reported sensitivity of 0.71 and specificity of 0.93. Two years later Amundson *et al* (2014) also investigating lung cancer reported sensitivities ranging from 0.62 to 0.65 and specificities from 0.08 to 0.16. Moreover there are inconsistencies in the methodologies used across the field (Moser and McCulloch, 2010). Here we aim to discuss both the advantages of canine detection and the shortcomings that need to be addressed such as sample collection methods and study design.

Method

*Our research*

Google scholar and pub med search engines were utilised for this paper. A number of combinations of keywords were used. The keywords were:- canine, dog, olfactory, smell, sniff, scent, detection, diagnosis, cancer, training, methods, disease, breath, urine, faeces, stool, bladder, breast, colorectal, lung, melanoma, prostate, ovarian, analysis, gas chromatography, mass spectrometry and volatile organic compounds (VOCs). In addition papers were also found via manual search, for example in the references of other relevant papers.

*Typical canine training and testing methodology*

Canines are trained using a reward-based system, usually food or a toy (McCulloch *et al*, 2006, Sonoda *et al*, 2011). The overall training duration and intensity of training varies a between studies. For instance Cornu *et al* (2011) spent 5 days a week over a 16 month period on training; whist Sonoda *et al* (2011) conducted their study over an 8 month period some 3 years after training began. Over the three year training period the canine was conditioned to detect a range of cancers including (but not exclusively): gastric cancer, lung cancer, breast cancer and colorectal cancer; all training was performed with breath samples, controls (n=500) were obtained by internet advertisement. It is worth noting the amount of cancerous samples used over this timeframe was not stated and nor was the frequency with which training sessions took place. A number of the reviewed studies do not report the duration of training (such as Ehmann *et al* (2012) and McCulloch *et al* (2006)). Training is often continued throughout the study duration as in Amundson *et al*, 2014 who maintained two training sessions per week, following low initial accuracy this training increased to four sessions per week. The test procedure usually takes place in a bare, familiar room (fewer distractions) with/without an observer present. These rooms usually (though not exclusively) share characteristics such as vinyl tiling, fluorescent lighting, no climate control, and some natural light (McColloch *et al*, 2006 and Sonoda *et al*, 2011). Cancerous samples and control samples are normally arranged in containers on the floor or in a carousel; canines (and their handlers) then enter the room and sniff each one in turn. It is noteworthy that of the reviewed literature the rational behind selection of the training and testing methodology employed (e.g. number of samples presented and use of carousel versus containers) is often not alluded to. The number of samples presented to the canine at any one time also varies between studies. Of the 14 papers reviewed the lowest number of samples in a test was 5 (4 control and one cancerous) (Rudnicka *et al*, 2014) and the highest 10 (8 control and 2 cancerous) (Horvath *et al,* 2010); though only 2 of the studies presented the canines with more than 6 samples at one time. Canines indicate (often by sitting in front of it (Ehmann *et al*, 2012)) which sample they believe has a differing odour pattern. A double blind study design was the most common amongst the studies reviewed (10 out of 13), 1 conducted both single and double blind tests; while 1 only conducted a single blind study. 1 paper did not indicate the study design employed (see Table 1.1). The double blind design ensures there are no conscious or subconscious indicators by the handler or laboratory assist (if used) to assist the canine performance (McCulloch *et al*, 2006).

Results and discussion

*Bladder cancer*

An early pilot study by Willis *et al*. (2004) trained 6 canines (of different breeds, not specified) over 7 months to distinguish between bladder cancer and control urine. 36 Patients with bladder cancer and 108 controls (either healthy or with benign disease) were used, 9 Cancer samples and 54 control samples were used for formal testing, with the remaining 27 patient samples and 54 controls used during the training period. Air-dried urine and liquid urine (previously frozen at -40$^o$C for up to 5 months and thawed) were used in separate tests. The canines had an overall mean success rate of 41% for dried and liquid urine tests combined, performing significantly better than chance (14% expected, based on 1 in 7 random chance of success as 1 cancerous and 6 control samples were used per test) in single blind tests. Dried urine sample testing could only achieve a 22% success rate compared to the 50% for liquid urine. There is no indication that the same canine was used for both a liquid and dried sample, this step could have aided in drawing possible conclusions on the handling and storage of urine samples for volatile organic compound (VOC) analysis. In each test one cancerous sample and six controls were used, one control for each test was age matched (some tests had two age matched controls); the age matched controls also had a non-cancerous urological pathology as did two non-age matched controls, the remaining control samples were from healthy participants. Control samples were sex matched to prevent hormonal interference.

Willis *et al*. (2011) later trained 4 canines of unspecified breeds to distinguish between bladder cancer urine and control urine. Urine samples were refrigerated within 45 minutes of collection, and then frozen at -80$^o$C until needed (up to six months). Using a double blind study, a sensitivity of 0.73 was obtained (from the best canine), though the group as a whole only obtained a sensitivity of 0.64. Specificity was reportedly 0.92 when cancerous urine was compared to urine from young, healthy controls; but a decrease in performance was seen when control urine was taken from elderly individuals with other benign urological diseases (specificity down to 0.56). This wide variation is suggestive that other alterations to urine odour patterns, due to age and other urological conditions can greatly reduce the efficiency of the canine ability to recognise specific cancer odour patterns. This highlights the need to select appropriate controls, this problem is not unique to canine studies a number analytical models also note drops in performance when age and symptom matched controls are introduced. The above results were generated from 210 participants, these were divided into 4 groups, control group 1 (n=61) were all younger than 33 and were healthy, control group 2 (n=65) had any non-cancerous and non-urological disease though did show abnormal urine, control group 3 (n=54) were a range of ages with benign urological diseases e.g. benign prostatic hyperplasia. 119 of the total participants were female, the remaining 91 were male. The remaining 30 participants all had confirmed transitional cell carcinoma. This paper does go into considerable detail of the participants involved and the clinical findings of each; it is not a common feature amongst the papers reviewed, to report in such detail on the controls samples used. None of the samples (patient or controls) were used during the training phase, the sample collection and storage of samples used in training was not commented upon.

Some work has been carried out to use sensors in an attempt to identify the VOC patterns associated with bladder cancer. Weber *et al,* (2011) used a gas sensor array in conjunction with pattern recognition software, to identify patients with bladder cancer. Although a small sample size was obtained the figures suggest further merit to the hypothesis that VOC differences can be associated with bladder cancer. In this instance 30 patient samples with new or recurrent bladder carcinoma were used. The control samples were comprised of three groups: 1) 20 healthy individuals, 2) 20 any non-cancer diseased patients, and 3) 19 non-cancer urological diseased patients. Urine samples were refrigerated immediately following collection then frozen at -80$^{o}$C; when needed these were then thawed and incubated at 38$^{0}$C. This same procedure of utilising urine samples is also commonplace amongst canine studies. This team found despite the presence of other non-cancerous urological disease in the controls, a sensitivity of 60% and specificity of 67%. Another group, Khalid *et al,* 2013 using a gas chromatography (GC) - sensor system with a statistical model, in a similar pilot study also had significant success in identifying VOC patterns in patients with bladder cancer. This modestly sized study of 98 participants, 24 with confirmed bladder cancer, and 74 control participants, had all undergone assessment for urological symptoms (e.g. haematuria) yielding a 95.8% leave-one-out cross-validation prediction value. Urine samples were stored at -20$^{o}$C and defrosted by immersion in an 80$^{o}$C water bath before the headspace was sampled with a Hamilton gas syringe. Whilst neither of these approaches identified the potential biomarker compounds responsible for the observed differences they indicate further investigation into odour patterns in the urine of bladder cancer patients is warranted.

*Breast Cancer*

Breast cancer is the commonest cancer among women in the UK (Cancer Research UK, 2014). Routine screening is already in place for breast cancer in the form of mammograms, which have greatly improved prognosis by encouraging early detection. However, mammograms are only offered to women over the age of 50 and therefore do not improve early diagnosis in younger women, and there has been some criticism that mammograms may lead to over diagnosis and unnecessary treatment (Kalager *et al*, 2012).

McCulloch *et al*. (2006) studied canines for both breast and lung cancer detection. They trained 5 canines (3 Labrador Retrievers and 2 Portuguese Water Dogs aged 18 months – 7 years) over an unspecified duration to detect breast cancer on exhaled breath. Breath samples were collected from 31 confirmed breast cancer patients (as soon as possible after biopsy, but prior to treatment commencement) and 83 volunteers with no previous cancer history (not age/symptom matched). It is worth noting that no details as to the time between biopsy and breath sample collection are reported.

There is no suggestion that the 83 control participants were suffering from any other non-cancer pathology however this is not clearly stated. These samples were collected in polypropylene sampling tubes obtained from Defencetek, South Africa and stored at room temperature for up to 60 days prior to analysis. A search into validation of these polypropylene sampling tubes revealed no information as to how long the sample can remain viable or about the best conditions to store them; since this appears to be a common technique throughout multiple canine studies validation of this technique would go some way to improving the reliability of such studies. However since at this stage the VOCs captured are unknown, developing a technique to validate this method of breath sampling is at best problematic. Furthermore this paper fails to report on how long the control samples were stored for and under what conditions; although a Fischer 2-sided exact test revealed "*There was no statistically significant difference between patients and controls in the time from breath sampling to testing*" (McCulloch *et al*, 2006). The reported result of P=0.52 for the Fischer 2-sided exact test is the only information provided to substantiate this quote, there is no reported method in this study for this calculation or explanation of how this conclusion was reached. Only 6 breast cancer patients (with 17 controls) were used in testing; these yielded a sensitivity of 0.88 and specificity of 0.98 using a double blind method; both were found to be similar across all stages of disease. All breath samples sniffed during formal testing were from completely different subjects not previously encountered by the canines during training. Willis *et al,* 2011 demonstrated symptom matched controls can have a negative affect on canine performance. The absence of symptom matched controls by McCulloch *et al*, (2006) may therefore have contributed to the seeming success of this study (which is noted by the author).

Breath VOC analysis with thermal desorption gas chromatography- mass spectrometry (GC-MS) by Phillips *et al,* (2006) identified: "*2-propanol, 2,3-dihydro-1-phenyl-4(1H)-quinazolinone, 1-phenyl-ethanone, heptanal, and isopropyl myristate*" (Phillips *et al,* 2006) as potential biomarkers of breast cancer. In the discussion the author states the biological basis for production of these compounds remains elusive. These compounds were selected as potential biomarkers based on a subtraction chromatogram; VOC abundance in the ambient air was subtracted from the VOCs in the breath sample. Analysis was carried out using a fuzzy logic statistical program. While this is encouraging the paper does use only a small sample of 51 asymptomatic patients with abnormal mammogram results (cancer diagnosis later confirmed through biopsy) and 42 age-matched healthy controls. This test predicted breast cancer with 93.8% sensitivity and 84.6% specificity. The follow up study (Phillips *et al*, 2010) paid considerably more detail to both study design and statistical analysis. This study used 54 patients with confirmed breast cancer (samples taken prior to treatment); and 204 cancer free controls. 1 Litre of breath and 1 litre of ambient air were collected on to sorbent traps; this allowed correction for any VOCs in the ambient air. The sensitivity (87.5%) and specificity (79.7%) were lower than that of the previous study, 30 volatiles were identified as potential markers of breast cancer, including: - cyclopropane ethylidene, tridecane, tetradecane, cyclotetrasiloxane octamethyl, and D-Limonene. It worth noting the 5 potential biomarkers noted by this group in their pilot study (Phillips *et al,* 2006) was not mentioned in this later work. This paper postulates the activation of cytochrome P450 in breast cancer provides the basis for the variation in breath VOCs, the authors hypothesis suggests a high-risk phenotype of cytochrome P450 increases the products associated with oxidative stress e.g. alkanes; thus increasing exhaled VOCs.

*Colorectal cancer*

Sonoda *et al*. (2011) used exhaled breath and stool samples for canine discrimination between patients with and without colorectal cancer. The canine was an 8-year-old black Labrador. Stool samples were obtained by suction during colonoscopy. The sensitivity using stool samples was 0.97 and the specificity was 0.99. It is not stated if the same participants were used for both stool and breath collection or if a different cohort was used for each sample type. Breath collection was performed by the participants exhaling approximately 100-200 millilitres of air into a breath collection bag obtained from Otsuka pharmaceutical company, Japan. These breath bags were placed into a zip-lock bag and stored at 4$^{o}$C; a full exhalation (from beginning to end) was obtained. In a number of other studies, it is usually only the end volume of breath collected to avoid collecting breath from the alveolar 'dead space', the authors fail to allude why this was not the case in this instance. The sensitivity using breath samples was 0.91 and the specificity was 0.99. All control stool samples were taken from patients undergoing investigative colonoscopy, so controls were symptom-matched. During testing, samples not encountered previously by the canines were used, furthermore each test sample and control sample were only used once. For breath testing 33 patient samples and 132 controls were used; for stool testing 37 patient samples with 148 controls were used. There is no information given as to how many samples were used for training, or how, when, and where the training samples were obtained. Accuracy was found to be high even for early-stage cancers, providing strong evidence for the possibility both breath and stool samples emit VOC patterns specific to colorectal cancer. This study performed the test trials during the winter/ spring months; the authors claim this is due to the canines lack of concentration during the warmer weather; this raises important questions as to the reliability of canines and affects of numerous external influences on canine behaviour though it should be noted the authors do not provide any evidence to support this claim.

A 2013, study had success in identifying colorectal cancer with an accuracy of 75% using VOCs from exhaled breath, via GC-MS (Altomare *et al,* 2013). In this study breath was collected in Tedlar® bags from fasted participants, the participants also had to acclimatise to the room the breath was collected in for at least 10 minutes to create an equilibrium between ambient air and lung air; additionally the sampling procedure had undergone some validation work in a previous study (Dragonieri *et al,* 2007). Background VOCs from Tedlar® bag production such as N,N-dimethylacetamide were also considered before analysis took place. Sample collection details such as participant fasted status, ambient air conditions, and validation of technique are rarely reported in canine studies. Altomare *et al,* 2013, also showed that while no-one biomarker stood out; 58 VOCs were initially identified these were then whittled down by eliminating: background VOC from the Tedlar$^{®}$ bags, and infrequently occurring VOCs. The remaining VOCs were narrowed down further with a box plot. 15 VOCs remained as potential biomarkers, these included: - decanal, nonanal, 2-methylbutane, cyclohexane, 4-methyloctane. These hydrocarbons and aldehydes could be due to oxidative stress.

*Lung cancer*

Lung cancer is the commonest cancer in the UK. Currently there are no suitable options for the routine screening of lung cancer; therefore the development of an effective technique suitable for mass screening is greatly needed.

McCulloch *et al.* (2006) trained 5 ordinary household canines (3 Labrador Retrievers and 2 Portuguese Water Dogs aged 18 months – 7 years) to assess lung cancer detection on exhaled breath. Breath samples were collected from 55 confirmed lung cancer patients (as soon as possible after biopsy) and 83 volunteers with no previous cancer history (not age/symptom matched). As mentioned previously in this paper there is no suggestion that the 83 control participants were suffering from any other non-cancer pathology however this is not clearly stated. Moreover since there is no reported time frame between biopsy and breath collection it possible the breath VOC alterations may be a result of residual anaesthetic effects. Harrison *et al* (2003) demonstrated that it was possible to see metabolites of the anaesthetic propofol using proton transfer reaction mass spectrometry on exhaled breath. Thus in the McCulloch *et al* (2006) study, since the control participants had been anaesthetised it is possible the canines were distinguishing between anaesthetised and non-anaesthetised patients. 28 patient samples were used for the double blind testing phase of the trial along with 17 controls. A sensitivity and specificity of 0.99 was achieved, using a double blind method. The sensitivity and specificity was found to be similar across all stages of disease (this study has been discussed in this paper see breast cancer).

Ehmann *et al.* (2012) trained four canines (two German Shepherds, an Australian Shepherd and a Golden Retriever) to detect lung cancer on exhaled breath. It was a large study with 220 participants, either healthy (n = 110), with histologically confirmed lung cancer (n = 60), or with chronic obstructive pulmonary disease (COPD) (n = 50). An overall mean accuracy of 73% and an overall sensitivity of 0.71 and specificity of 0.93 were obtained. The inclusion of a COPD group allowed the authors to determine that in this study the canines appeared to be able to distinguish between lung cancer and COPD patients; this suggests each disease has its own unique odour. Moreover the lung cancer odour pattern was also detectable despite smoking status or diet of the patients. These results were obtained using a double blind method, and different breath samples were used during training and formal testing. These numbers at first inspection appear encouraging though it should be noted there was a considerable variability between individual canines with accuracy ranging from 68% to 84%. Unlike many other studies of this kind the authors have presented figures for PPV and NPV of 75% and 93% respectively. This study does contain significant detail regarding the methods employed however it does mention two separate sample collection sites; though there is no mention which samples were collected at which site. This study attempts to display canine odour recognition as a robust method and thus contains no participant restrictions such as diet or smoking status.

More recently Amundson *et al.* (2014) tested the olfactory ability of 4 canines a Belgian Shepard, Boarder collie, Hard hair dachshund, and Rottweiler to detect lung cancer using exhaled breath and urine. Samples were taken from 93 patients admitted to hospital with suspected lung cancer; none of the participants were healthy, 59 were diagnosed with lung cancer. Following low accuracy on an interim analysis, a period of "*intensive training*" (Amundson *et al*, 2014) was employed before a second block of testing was carried out; this simply involved doubling the weekly training sessions. The breath test samples were divided into two groups, the first group of 46 participant samples had an overall sensitivity of 64.7% (ranging from 67.6% to 61.8% for individual canines); and overall specificity of 8.3% (16.7% to 8.3% for individual canines). A period of intensive training followed before testing the next group of 40 patients. This second test yielded an overall sensitivity of 56% and specificity of 33.3% (ranges from 56% to 76% sensitivity for individual canines and 33.3% to 53.3% specificity). 77 urine samples were available from the same set of patients all samples were tested twice, again, with a period of intensive training between the first and second tests. The first test overall sensitivity and specificity were 73.6% and 25% respectively and 64.2% and 29.2% respectively for the second test. Individual canine sensitivity ranged from 49.1%- 71.7% across the two tests and specificity ranged 20.8%-41.7%.

Rudnicka *et al*. (2014) investigated canine olfactory detection of lung cancer on exhaled breath alongside GC/MS analysis. Based on our search this is one of few instances in which analytical techniques such as GCMS have been employed in conjunction with canine testing. This approach would seem a logical step in identifying the distinct odour pattern the canines appear to be detecting. The GCMS analysis was used in conjunction with solid-phase micro extraction (SPME) pre-concentration. Two male German Shepherds of unknown ages were used. This study used breath collected from 108 patients with confirmed lung cancer, 121 healthy controls and 24 controls with other respiratory disease. The canine sensitivity was reported at 0.86 and specificity 0.72 using a double blind method. Again, it is unclear whether any samples used during training were re-used during testing. GC-MS analysis yielded a sensitivity of 0.74 and a specificity of 0.73. The authors note several compounds found in significantly higher concentrations in samples from cancer patients than those from healthy controls; acetone, isoprene, ethanol, 1-propanol, 2-propanol, hexanal and dimethyl sulphide the authors deemed most noteworthy. Despite the promising canine figures again we see large variation between the two individual canines. There appears to be significant statistical analysis with positive and negative predictive rates being calculated.

*Melanoma*

Pickel *et al*. (2004) found two canines (a 4 year old Standard Schnauzer and 6 year old Golden Retriever) were able to differentiate melanoma (skin cancer) from controls, although sensitivity and specificity were not stated. This group conducted three separate trials, initially box trials were conducted where by the malignant sample was place in one of 10 wells, and the remaining wells were filled with materials such as latex gloves or gauze in order to distract the canine. The amount of filled wells varied from trial to trial and empty wells were not considered during results analysis. Canine A performed 16 box trials and canine B 11 box trials. It is not stated if the same malignant samples were used for both canines. In each case both canines performed with a 100% success rate. This prompted the beginning of the next round of testing in which malignant skin biopsies were placed on the bodies of volunteers in sticking plasters. The placement of the plaster (sample) was varied each time, each dog was tasked with locating 5 samples each and again both gave 100% success rate. Both canines were then instructed to localize the melanoma source on 7 live patients, canine A correctly localized melanoma on six of the patients. Canine B sniffed 4 of the same patients (it is not stated why the remaining 3 weren't tested on canine B), and correctly localised melanoma in three of the patients. The design of this study is questionable; melanoma tissue was placed on volunteers, which may not be relevant to live tumour detection; the authors suggest the VOCs emitted maybe from the underside of the removed tissue. Moreover tissue samples are known to deteriorate after a matter of hours and thus the canines maybe detecting VOCs associated with necrotising tissue. Dogs were also taken directly to a patient to sniff, which has been avoided by other studies. The results of this study are interesting but do not lend themselves as easily to potential screening applications. It is also unclear whether any test samples had previously been encountered by dogs during training. As is consistent with canine studies we again see variability in performance between individual canines.

As seen with other cancers we again find some evidence that melanoma cells do exhibit an altered VOC signature compared to healthy cells. A study, once again, using SPME GCMS this time coupled with a single-stranded DNA coated nanotube sensor (DNACNT) was able to elucidate some VOC alterations in melanoma cells when compared to that of healthy melanocytes (Kwak *et al*, 2013). Although numerous VOCs differed between the two cell lines the team was able to identify two compounds (dimethyldisulfide and dimethyltrisulfide) unique to melanoma cells (Kwak *et al,* 2013).

*Ovarian cancer*

Ovarian cancer is the fifth most common cancer among women (Cancer Research UK, 2014); however it has a very poor prognosis, mainly owing to late detection (symptoms may be very similar to other conditions). Therefore developing an adequate method of screening is of great importance to affected individuals.

Horvath *et al*. (2008) trained a 4 year old Riesenschnauzer to detect ovarian cancer from tissue samples. 31 different tumour samples were obtained and each one divided into 10-30 individual samples these were stored in a -80$^o$C freezer and thawed for 15-30 minutes before use; all tissue samples were treated in the same manor (including controls). A number of different tests were conducted; single blind tests were conducted to determine if the canines could distinguish between ovarian cancer and healthy tissue. In this test tumour samples had been used during the training process (4 patient samples), the control samples were comprised of fat and abdominal muscle samples (unspecified quantity); in this case the canine achieved sensitivity and specificity of 1.0. In the following single blind test ovarian carcinoma tissue (from the same 5 patients as previous test) was used along side an unspecified number of non-ovarian gynaecological carcinomas. In this instance sensitivity was again 1.0 and specificity of 0.91 (the result of three false positives). For the double blind testing 20 new samples were used and had not been previously exposed to the canine; the controls (again unspecified amount) comprised of healthy muscle, fat and small bowel samples. In this instance the canine again achieved a sensitivity of 1.0 and a specificity of 0.975. The tumour sample and control samples were kept in separate rooms, to avoid odour contamination; however this presents a potential flaw as it is possible the samples were contaminated with the ambient air and thus the canine may have been distinguishing the room the sample was prepared in. Furthermore the controls used were not obtained from a gynaecological source but were in fact other tissue types such as small bowel, which may already containing differing odour patterns than healthy gynaecological samples.

Tissue biopsies are invasive and unsuitable for routine screening; blood sample collection, however, is less invasive and a more promising option. Horvath *et al.* (2010) were able to use 2 Giant Schnauzers (aged 7 and 3) to detect ovarian cancer from blood samples, with a sensitivity of 1.0 and specificity of 0.98. Blood was collected from ovarian cancer patients before primary cancer surgery, and control blood was collected from healthy (mainly young) females and some male blood samples were used as controls (the authors report this did not effect canine performance). These samples were spun in a centrifuge, the collected plasma retained and frozen at -80 $^0$C until required for testing. Different samples were used for training and testing; it is unclear how many patient and control samples were collected. It should also be noted that some of the controls used alongside the patient plasma were of differing tissue such as abdominal fat and other gynaecological tumour tissues e.g. cervical carcinoma. This study shows very little variation between the different canines, and seems to address a number of causes for concern in the previous study, Hovarth *et al*, 2008. The discussion raises the interesting point that the canine nose can only detect odour molecules whilst analytical techniques such as the electronic nose used in previous work by the team will potentially be able to detect non-odorous compounds such as methane. Conversely it is also possible the canine nose is more sensitive to certain compounds that may be beyond the detection range of an electronic nose.

A recent pilot study however has used GCMS and a separate nanoarray of sensors to analyse breath samples of 182 participants (48 with ovarian cancer) (Amal *et al,* 2015). Although the results were not as good as those seen in the discussed canine studies the team achieved an accuracy, sensitivity and specificity of 71% and were able to identify several potential biomarkers including 2-butanone and hexadecane. The nanoarray technique performed considerably better with accuracy, sensitivity, and specificity of 0.89, 0.79, and 1.0 respectively. The obtained signals from the sensor array were used to build diagnostic classifiers with discriminant factor analysis. Using this method the nanoarray was able distinguish early from advanced cases of ovarian cancer (Amal *et al*, 2015).

*Prostate cancer*

Prostate cancer is the most common cancer among men in the UK (Cancer Research UK, 2014). The current standard screening method for prostate cancer is the prostate specific antigen (PSA) test. However it is only indicatory, not diagnostic. PSA levels can be raised in other benign conditions, or may not be raised at all in some men with prostate cancer leading to false negatives (Catalona *et al*, 1991). Therefore there is a real need for an alternative test.

Cornu *et al*. (2011) trained a Belgian Malinois to distinguish between the urine of 33 men with biopsy confirmed prostate cancer and urine from 33 controls, with a sensitivity and specificity of 0.91, generated from new samples in a double blind manner. A total of 108 urine samples were collected; 26 cancerous samples and 16 controls were used during the training phase and the remaining samples used for testing. All participants had been referred with raised PSA levels and/or abnormal digital rectal examinations so were age/symptom matched, and urine was collected (and frozen) upon the first urologist consultation. After biopsy they were placed in the cancer or control category. Separate samples from different individuals were used for training (26 cancer samples and 16 control samples). It was hypothesised that the biochemical difference between the samples may be attributed to the presence of significantly increased levels of sarcosine in cancerous urine, as found by Sreekumar *et al*. (2009), however subsequent studies have failed to replicate these findings. The use of only one canine does limit this study (as mentioned by the author); furthermore samples were frozen at -4$^o$C, for an unspecified time period compared to the more usual standard of -80C. This paper does describe in some detail the limitations of this kind of study, for instance the small sample sizes making real conclusions hard to define. The authors also raise the point the process of training and testing the canines is both expensive and lengthy, greatly limiting the potential for clinical application.

A later study by Elliker *et al*. (2014) found that although two canines (a 9 year old Labrador and a 3 year old Border collie) appeared to have learned to select for prostate cancer in urine samples during training, they did not select for prostate cancer when presented with new samples. The authors attributed lack of success to canine memory of control samples, which is of concern. 31 Patients with small to metastatic prostate cancer tumours and 93 controls (either healthy or with benign prostatic hyperplasia, information regarding the controls used during testing is sparse) took part in this study, which used a double blind design. 50 cancer samples and 67 control samples were used during training, and 31 cancer samples and 93 control samples were used during testing. Crucially this paper makes a unique point of noting that all samples (control and patient) were collected in the same place. This process aims to eliminate contamination of odours associated with the ambient air of clinical settings, though ideally to further enhance this process all samples should be collected at the same time. The authors also encounter difficulties in the variability of individual canines, which is reported in the majority of studies. Moreover the temperament of the canines meant that only three of the ten canines selected progressed past the first training stage due to either excitability or inability to motivate the canine to task. This paper highlights that in a number studies it is not stated whether during the testing phase (double-blind or otherwise) completely new patient/subject samples are used so that the dog has had no previous exposure to the samples in training. Once a dog has received training on a specific individual the dog will be conditioned to this person's odour as a negative or positive sample. This means that if it is not stated that the patients and samples are completely new to the dog the results are likely to show an inflated sensitivity/specificity. Only results from papers that clearly state that patients and samples were completely new to the dog during testing can be considered reliable.

A very recent study by Taverna *et al* (2015) has provided very promising results. 2 3-year-old German shepherd canines were used to identify prostate cancer from urine. These canines had previously been trained in explosives detection; in this instance the canines underwent full time operant conditioning training to differentiate prostate cancer patients from controls. A total of 902 participants were recruited for this study. The cancer group consisted of 362 males split into 5 groups: group 1 had been treated with surgery (n=180), group 2 had increased PSA and histological diagnosis of prostate cancer following biopsy (n=120), group 3 had prostate cancer detected incidentally (n=22), group 4 had metastatic prostate cancer or were receiving hormone therapy (n=29), and group 5 had primary prostate cancer and another tumour (n=11). Similarly the control group of 540 was also split into sub groups. 122 of the control participants were female, 50 of who were healthy and 72 of whom had non-neoplastic conditions. The use of controls of the opposite sex has been avoided in other studies as canines many be able to detect hormonal differences to differentiate male from female participants (Willis *et al*. 2004). If the canines are easily able to differentiate male from female samples this may this may partially increase the chance of the canine successfully identifying the correct sample. The authors state the use of female controls during the early testing phase prevented the canines from being confused by prostrate specific VOCs. The authors do go on to perform statistical analysis, which discounts the female controls. The remaining controls were split into 4 other groups including: healthy males, males 45+ years old with urological and/or systemic disease, urinary obstruction due to benign prostatic hyperplasia, and a group with low PSA and negative family history of prostate cancer with negative results of digital rectal exam and non-prostatic cancer. All urine samples were stored at -20$^o$C and controls and patient samples were kept separately to avoid contamination. 200 patient samples and 230 controls were used for training and were not reused for testing; moreover all testing was performed in a double blind manor. The authors present the findings of the 2 canines separately and do not combine the results as many other papers do; this means the variability between the canines is transparent (and very low in this instance) and prevents inflation of overall results by averaging. Canine 1 achieved an overall sensitivity of 1.0 and specificity of 0.987 (which dropped to 0.983 once excluding female controls and 0.98 when considering only controls of 45 year old plus males). Canine 2 was only slightly inferior with an overall sensitivity of 0.986 and specificity of 0.976 (which dropped to 0.969 once excluding female controls and 0.964 when considering only controls of 45 year old plus males).

As with ovarian cancer we again find sparse evidence to elucidate what VOCs are present in cases of prostate cancer. Smith *et al,* (2010) had some success in identifying VOC alterations in patients with prostate cancer. Additionally formaldehyde has also been postulated as a possible biomarker, though this is now somewhat outdated and doesn't appear to have been brought to fruition (Spanel *et al,* 1999).

*Table 1.1 Comparison of canine studies*

| | Cancer type | Sample type | Cancer stage | Quoted Sensitivities/ specificities | No. of canines/ breeds | Participants/ participants with cancer | Blinding | Same samples used for training and tests |
|---|---|---|---|---|---|---|---|---|
| **Amunsdon et al. (2014)** | Lung | Breath | Not stated | 0.618-0.676/ 0.083-0.167 | Belgian Shepard, Boarder collie, Hard hair dachshund, and Rottweiler | Not stated/ 93 | Double blind | Not Stated |
| **Cornu et al. (2011)** | Prostate | Urine | Not stated | 0.91/ 0.91 | 1/ Belgian Malinois | 66/ 33 | Double blind | No |
| **Ehmann et al. (2012)** | Lung | Breath | Not stated | 0.71/ 0.93 | 4/ 2 German Shepherds, Australian Shepherd, Golden Retriever | 220/ not stated | Double blind | No |
| **Elliker et al. (2014)** | Prostate | Urine | Small to metastatic tumours | Not stated | 1/ Labrador, Border Collie | 124/ 31 | Double blind | No |
| **Horvath et al. (2008)** | Ovarian | Tissue | Borderline to metastatic tumours | 1.0/ 0.98 | 1/ Riesenschnauzer | Not stated/31 | Single and double blind | No |
| **Horvath et al. (2010)** | Ovarian | Tissue | Borderline to metastatic tumours | 1.0/ 0.98 | 2/ Giant Schnauzers | Not stated | Double blind | No |
| **McCulloch et al. (2006)** | Lung and breast | Breath | Range of disease stages | 0.99/ 0.99 for lung cancer 0.88/ 0.98 for breast cancer | 5/ 3 Labrador Retrievers, 2 Portuguese Waterdogs | 169/ 86 | Double blind | No (breast) No (Lung) |
| **Pickel et al. (2004)** | Melanoma | Tissue | Not stated | Not stated | 2/ Standard Schnauzer, Golden Retriever | Not stated/ 7 | Double blind | Not Stated |
| **Rudnicka et al. (2014)** | Lung | Breath | Not stated | 0.86/ 0.72 | 2/ German Shepherds | 253/ 108 | Double blind | Not Stated |
| **Sonoda et al. (2011)** | Colorectal | Breath and stools | Range of stages | 0.91/ 0.99 for breath 0.97/ 0.99 for stools | 1 /Labrador | 350/ 70 | Not stated | No |
| **Taverna et al. (2015)** | Prostate | Urine | Very early to metastatic tumours | Canine 1 1.00/ 0.987- 0.98 Canine 2 0.986/ 0.976- 0.964 | 2/ German Shepherds | 540/ 362 | Double blind | No |
| **Willis et al. (2004)** | Bladder | Urine | New or recurrent carcinoma | 41% success rate | 6/ not stated | 144/ 36 | Single blind | No |
| **Willis et al. (2011)** | Bladder | Urine | Not stated | 0.64/ 0.56- 0.92 | 4/ not stated | 210/ 30 | Double blind | No |

*Overview*

Canines are reported to have been assessed for their ability to differentiate cancers from healthy controls, and are summarised in Table 1.1, along with the values obtained for sensitivity and specificity. The most notable feature of Table 1.1 is the amount of "not stated" values; omissions of key facts, particularly in methodology, calls into question the validity of the produced statistical values. Although the levels of detection have, in some cases been remarkably high and superior to the current commercial tests (e.g. PSA test for prostate cancer).

A review paper by Horvath *et al.* (2009) reported on several analytical methods of detecting lung cancer biomarkers including SPME-GC, GCMS with sorbent trap, electronic noses, and canine olfaction, and found that canine olfaction achieved sensitivity and specificity superior to any other method (0.99, found by McCulloch *et al.*, 2006). Though there are significant limitations and questions to be raised by the McCulloch *et al,* 2006 study which have been discussed (see section headed *breast cancer*). Furthermore since all the techniques discussed in the Horvath *et al.* (2009) review are highly variable (in terms of sample collection and processing, and pre-concentrating methods) comparison of these methods against one another is inappropriate.

The cancer diagnoses in canine studies have typically been confirmed by gold standard biopsy (followed by microscopy). It should be considered though that canine detection is very much a positive/ negative test as communication of quantification of odour patterns from canine to human would be highly subjective (if at all possible). Thus the canine "diagnosis" could not be considered for prognosis or treatment response measurement, both of which are critical to an effective biomarker. Moreover studies haven't been large enough to undertake rigorous statistical analyses with the majority of studies simply quoting sensitivities and specificities; while these figures are of great importance to a diagnostic test they are just two of multiple factors contributing to the effectiveness of a test. As Cornu *et al* (2011) remarked the process of training and testing the canines are both time consuming and costly. That said, once a canine has been trained the testing process is very rapid; once presented with a sample(s) the canines in the above studies are able to make a decision within seconds as noted by Cornu *et al* (2011) (quoted a mean time of 30 seconds). For comparison the run time for a sample with the GCMS method used by Rudnicka *et al* (2014) took in excess of thirty minutes; this does not include time to condition the equipment before sample collection and pre-concentrating can take place.

Clearly for a good robust canine study, age-matched and symptom-matched controls are required. These may be patients with benign illness, whose symptoms triggered them to seek clinical investigation, e.g. Amundson *et al.* (2014) controls had been admitted to hospital with suspected lung cancer, and subsequently found to be negative for the disease. In the majority of the studies commented upon here a mix of healthy and symptom matched controls are used; these often include benign tumour samples. However, it is sometimes the case that only 'healthy' individuals have been used as controls. When using symptom matched controls it is not possible to know for certain if the individual is cancer free or subject to a false-negative test result. For example the study by Cornu *et al*. (2011); in which control samples were collected from patients that had abnormal levels of PSA and abnormal digital rectal examination (DRE); found during testing three of these controls were identified as positive by the canine all three patients had a second biopsy; one of these was confirmed as having prostate cancer. Studies, which have selected age-matched and symptom-matched controls, do show lower levels of sensitivity (e.g. Amundson *et al.,* 2014), indicating that the task becomes more difficult as the level of 'noise' increases, making the cancer signature harder to detect. It should be noted however this is often the case for more traditional analytical studies and is not a problem exclusive to canine detection.

A potentially confounding factor in studies using breath samples is the site at which samples are collected; ideally samples should be collected from individuals in similar environments at the same time of day, or preferably all samples should be taken in one session; though in reality this is unlikely to be feasible. For example, if diseased (cancer) samples are taken from patients in a hospital environment and control samples are not collected in a clinical setting, it is possible that hospital volatiles may be present and improve dogs' performance. The papers reviewed do not clearly state where samples were collected, which should be addressed in future studies. Ideally both patient samples and controls should be collected in a non-clinical setting to prevent clinical ambient VOCs contaminating the healthy controls. Since this scenario is often not possible, as some patients will have no option but to be in a clinical setting, control participants should therefore be invited to the same clinic as the patients (though this remains suboptimal since ambient VOCs also vary with time). Storage and collection of breath samples also have inherent complications. Most canine studies collect breath using oil coated polypropylene fibres in a plastic tube, which the participant then breath's over (usually 3 deep exhalations, as in Rudnicka *et al,* 2014). This means the quantity of air passing over the fibres will be highly variable between participants, particularly for use in lung cancer testing, it may in this instance be a lack of odour the canine is responding to. Furthermore the study by McCulloch *et al*, (2006) kept breath samples in a sealed plastic bag for up to 60 days without providing any validation work to ascertain the viability of the samples over time. Our own search into validation of this breath sampling technique also yielded no results despite this method being commonly used in canine studies and for forensic applications. The collection, handling and storage of 'hard' samples (blood, urine, and stool) tend to be standardised in that very soon after collection these samples are frozen at $-80^{o}C$ or in some cases $-20^{o}C$ until defrosting for use. Once thawed these samples are often warmed to body temperature ($37^{o}C$) and used within an hour.

Perhaps the most important issue faced by canine studies is the amount of variability between individual animals. This may contribute to the seemingly high success rates in studies only using one canine. The study by Ehmann *et al* (2012) highlights this point well with their canines demonstrating an accuracy varying between animals of 0.68- 0.84. By averaging out these results or only reporting the upper figures it is to see how some studies might show artificially high results. The reason for the variability has been commented upon in Cornu *et al* (2011), which cites "*olfactory-receptor* *polymorphisms*" (Cornu *et al*, 2011) as the cause. There is evidence to suggest the G- allele on the cOR9S13 gene can have a positive effect on binding affinity of some odour molecules to olfactory receptors (Lesniak *et al*, 2008). This points the way to suggest that genetic studies may pave the way for the efficient selection of canines in cancer odour studies. There is also the possibility to use this knowledge to develop biosensor technology, where by grown cells with these genes can be utilised for their binding affinities.

Furthermore, in order for canine detection to be clinically useful, studies will need to be done on the canine's performance over time. The studies reported in this paper do not give an accurate description of the time lapse between training and testing; though it is likely this is a short time period. Should the canine require constant training to maintain performance this will significantly increase the overall costs of testing and reduce the possible efficiency of testing sites. Canines will also have to be selected based on behaviour patterns as Elliker *et al*. (2014) eluded to; in this instance only three of the ten originally selected canines showed the necessary behaviour to progress past the first round of training. This again could significantly raise the overall costs associated with this type of testing.

An option to introduce the clinical utility of canines might be to collect samples in the clinic, and transport them to specialist screening centres, using carousel systems and stands, such as those used by APOPO rats for tuberculosis detection, or the RASCargO™ system, used for canine detection of explosives in cargo. Potentially thousands of samples could be screened in this way. Information could then be sent to the clinician to be added to the whole picture, before being discussed with the patient; importantly the canine does not diagnose but simply provides an additional test result. Though this approach will detract from the appealing rapidity of canine testing by adding significant time between sample collection and receiving a result; thus in order to make this a cost effective solution reliability of results must be significantly improved.

Recommendations for future canine studies

In order improve the reliability of reported findings in canine studies a number of key areas need to be addressed and implemented across the field.

*Sample collection and storage*

1. Breath sample storage media for use in canine studies (polypropylene fibres as in McCulloch *et al,* 2006) should undergo rigorous validation study. This should set out the most appropriate storage of sample tubes and set time limits the samples should be used by; in addition to being made readily available for reference.
2. All breath samples should ideally be collected in the same room; including controls as in Elliker *et al*. (2014). This process will go some way eliminate the potential contamination of the sample with ambient air volatiles. Though as mentioned this is still not perfect as ambient VOCs can alter over time.
3. Tissue, urine, and stool sampling have a well-defined collection and storage procedure. In this instance samples should be chilled immediately after collection to prevent degradation effects such as tissue necrosis and urine alterations. Storage over longer periods of time (as is often required for such studies) should ideally take place in -80$^{\circ}$C freezers. If samples are collected from patients under anaesthesia then conventional analysis should be undertaken to ensure no residual VOCs from the anaesthesia remain.

4. Once frozen samples are thawed and warmed to 37$^{\circ}$C they should ideally be used within 2 hours to prevent microbial growth from interfering with VOC emissions.
5. Whole blood samples should be collected in EDTA vacutainers to stabilise folate, stored at 4$^{\circ}$C and used within 7 days, if analysis cannot take place in this time frame again -80$^{\circ}$C freezer storage is required (Tolonen *et al*, 2002).
6. Detailed information on sample handling, storage and usage should be reported in all publications including reference material of validation studies.

*Participants*

1. Inclusion and exclusion criteria for both controls and participants should be reported.
2. Controls should be sex matched to prevent hormonal interference (Willis *et al.* 2004).
3. Controls should include symptom matched non-cancerous samples to ensure the canine is not simply differentiating between healthy and non-healthy.

*Training*

1. There should be an agreed upon best practise training methodology used across the field. It is worth noting all the studies in this report use broadly speaking the same training procedure.
2. Detailed reporting of all methodologies employed should be included in publications. Including detailed information regarding the collection, storage, and quantity of samples used over the training duration.

*Testing*

1. All testing procedures should be thoroughly detailed in publications.
2. Crucially samples used in training must not be used in testing; all new samples are required for all tests, both patient and control.
3. Double blind method of testing should take place to prevent experimental bias.

*Analysis*

1. Clinical utility analysis should take place including PPV and NPV in addition to sensitivity and specificity. Canine studies usually only use small sample sizes, thus a detailed analysis is impractical and of limited value.
2. Where an ROC curve is drawn the procedure with which the author derives this should be included in publication.

*Other points to consider*

1. Individual canine results and the affect the variability between individuals has on the overall accuracy of the study should be reported in detail.
2. The longer-term effectiveness of training should be assessed. Can the same training method be employed to maintain the canine's efficiency? Additionally a full cost effectiveness evaluation should be conducted in order to ascertain any potential commercial viability.
3. Where possible laboratory analysis (using appropriate method e.g. GC-MS) of the same samples should be done in conjunction with canine testing. In addition to validating the canine test this could also provide insight into potential biomarkers.

<u>Closing remarks</u>

Taken together the evidence presented by the work reviewed in this report does suggest there is a basis for the hypothesis that the canine olfactory sense (once trained) is sufficiently sensitive as to differentiate cancerous from non-cancerous samples. However the outlined concerns must be addressed before more substantial conclusions can be reached. Furthermore there needs to be larger studies performed that can under go the rigorous statistical tests necessary to validate the claims made. Given the variability between individual canines it is highly unlikely that canine detection could ever be used in a clinical setting. Moreover canine testing will not be able achieve the consistent levels of control compared to that of a laboratory investigation.

Although some of the experimental design in canine studies can be called into question the same criticism can also be levelled at many analytical studies. One major advantage of many canine studies is that they incorporate both a model building (training) and true validation stage, where new patients are assessed by the trained canines. Admittedly this is an area which could be made more transparent within the studies so it is possible to determine the numbers used for both training and validation (and the number of patients retested from within the training set). However, many analytical studies identifying biomarkers of disease do not validate the statistical models on any samples outside the original patient cohort.

The primary benefit of canine investigations is the encouragement that it gives to the volatile analysis community, to search for biomarker compounds or VOC patterns detectable by man-made volatile analysis systems; such as those reported.

Acknowledgements: Clare Guest from the non-profit Medical Detection Dogs, Milton Keynes, MK17 ONP, United Kingdom

References

Altomare, D. F., Di Lena, M., Porcelli, F., Trizio, L., Travaglio, E., Tutino, M., Dragonien, S., De Gennaro, G. (2013) Exhaled volatile organic compounds identify patients with colorectal cancer. *British journal of surgery*, 100 (1), pp. 144-150.

Amal, H., Shi, D. Y., Ionescu, R., Zhang, W., Hua, Q. L., Pan, Y. Y., Li, T., Hu, L., Haick, H. (2015) Assessment of ovarian cancer conditions from exhaled breath. *International Journal of Cancer*, 136 (6), E614-E622.

Amundson, T., Sundstrom, S., Buvik, T., Gederaas, O. A. and Haaverstad, R. (2014) Can dogs smell lung cancer? First study using exhaled breath and urine screening in unselected patients with suspected lung cancer. *Acta Oncologica.* 53 (3), pp. 307-315.

Bijland, L. R., Bomers, M. K. and Smulders, Y. M. (2013) Smelling the diagnosis A review on the use of scent in diagnosing disease. *The Netherlands Journal of Medicine.* 71 (6), pp. 300.

Bomers, M. K., van Agtmael, M. A., Luik, H.,van Veen, M. C., Vandenbroucke-Grauls, C. M. J. E. and mulders, Y. M. (2012) Using a dog's superior olfactory sensitivity to identify Clostridium difficile in stools and patients: proof of principle study. *British Medical Journal.* 345 (7888), pp. 7-8.

Bomers, M. K., van Agtmael, M. A., Luik, H., Vandenbroucke-Grauls, C. M., Smulders, Y. M. A. (2014) detection dog to identify patients with Clostridium difficile infection during a hospital outbreak. *Journal of Infection* 69 (5), pp. 456-61.

Cancer Research UK (2014) *Early Diagnosis: Support Research to Detect Cancer Early*. Available from: http://myprojects.cancerresearchuk.org/projects/early-diagnosis-support-research-detect-cancer-early [Accessed 13 November 2014].

Cancer Research UK (2014) *Cancer Stats: Key Facts.* Available from: http://www.cancerresearchuk.org/cancer-info/cancerstats/keyfacts/ [Accessed 14 November 2014].

Catalona, W. J., Smith, D. S., Ratliffe, T. L., Dodds, K. M., Coplen, D. E., Yuan, J. J. J., Petros, J. A. and Andriole, G. L. (1991) Measurement of prostate-specific antigen in serum as a screening test for prostate cancer. *The New England Journal of Medicine.* 324 (17), pp. 1156-1161.

Cornu, J., Cancel-Tassin, G., Odnet, V., Girardet, C. and Cussenot, O. (2011) Olfactory detection of prostate cancer by dogs sniffing urine: a step forward in early diagnosis. *European Urology.* 59 (2), pp. 197-201.

Dragonieri, S., Schot, R., Mertens, B. J., Le Cessie, S., Gauw, S. A., Spanevello, A., Resta, O., Willard, N. P., Vink, T. J., Rabe, K. F., Bel, E. H., Sterk, P. J. (2007) An electronic nose in the discrimination of patients with asthma and controls. *Journal of Allergy and Clinical Immunology*, 120 (4), 856-862.

Ehmann, R., Boedeker, E., Friedrich, U., Sagert, J., Dippon, J., Friedel G. and Walles T. (2012) Canine scent detection in the diagnosis of lung cancer: revisiting a puzzling phenomenon. *European Respiratory Journal*. 39 (3), pp. 669-676.

Elliker, K. R., Sommerville, B. A., Broom, D. M., Neal, D. E., Armstrong, S. and Williams, H. C. (2014) Key considerations for the experimental training and evaluation of cancer odour detection dogs: lessons learnt from a double-blind, controlled trial of prostate cancer detection. *BMC Urology.* 14 (1), pp. 22.

Furton, K. G., Myers, L. J. (2001) The Scientific Foundation and Efficiency of the use of Canines asd Chemical Detectors of Explosives. *Talanta* 54 (3), pp. 487-500

Harrison, G. R., Critchley, A. D. J., Mayhew, C. A., & Thompson, J. M. (2003) Real-time breath monitoring of propofol and its volatile metabolites during surgery using a novel mass spectrometric technique: a feasibility study. *British journal of anaesthesia*, 91 (6), pp. 797-799.

Horvath, G., Andersson, H. and Paulsson, G. (2010) Characteristic odour in the blood reveals ovarian cancer. *BMC Cancer*. 10 (1), pp. 643.

Horvath, G., Järverud, G. A. K., Järverud, S. and Horváth, I. (2008) Human ovarian carcinomas detected by specific odor. *Integrative Cancer Therapies.* 7 (2), pp. 76-80.

Horvath, G., Lazar, Z., Gyulai, N., Kollai, M. and Losonczy, G. (2009) Exhaled biomarkers in lung cancer. *European Respiratory Journal.* 34 (1), pp. 261-275.

Kalager, M., Adami, H. O., Bretthauer, M., & Tamimi, R. M. (2012) Overdiagnosis of invasive breast cancer due to mammography screening: results from the Norwegian screening program. *Annals of internal medicine*, 156 (7), pp. 491-499.

Khalid, T., White, P., Costello, B. D. L., Persad, R., Ewen, R., Johnson, E., Probert C., Ratcliffe, N. (2013). A pilot study combining a GC-sensor device with a statistical model for the identification of bladder cancer from urine headspace. *PloS one*, 8 (7), e69602.

Kwak, J., Gallagher, M., Ozdener, M. H., Wysocki, C. J., Goldsmith, B. R., Isamah, A., Faranda, A., Fakharzadeh, S.S., Herlyn, M., Johnson, A.T.C., Preti, G. (2013) Volatile biomarkers from human melanoma cells. *Journal of Chromatography B*, 931, 90-96.

Lesniak, A., Walczak, M., Jezierski, T., Sacharczuk, M., Gawkowski, M., Jaszczak, K. (2008) Canine olfactory receptor gene polymorphism and its relation to odor detection performance by sniffer dogs. *Journal of heredity*, 99 (5), pp. 518-527.21

Lippi, G., & Cervellin, G. (2012). Canine olfactory detection of cancer versus laboratory testing: myth or opportunity?. *Clinical Chemistry and Laboratory Medicine*, 50 (3), pp. 435-439.

McCulloch, M., Jezierski, T., Broffman, M., Hubbard, A., Turner, K. and Janecki, T. (2006) Diagnostic accuracy of canine scent detection in early and late-stage lung and breast cancers. *Integrative Cancer Therapies*. 5 (1), pp. 30-39.

McGuire, N. D., Ewen, R. J., de Lacy Costello, B., Garner, C. E. and Probert, C. S. J. (2014) Towards point of care testing for C. difficile infection by volatile profiling, using the combination of a short multi-capillary gas chromatography column with metal oxide sensor detection. *Measurement Science and Technology.* 25 (6), pp. 06510.

Moser, E. and McCulloch, M. (2010) Canine scent detection of human cancers: A review of methods and accuracy. *Journal of Veterinary Behaviour: Clinical Applications and Research*. 5 (3), pp. 145-152.

Phillips, M., Cataneo, R. N., Ditkoff, B. A., Fisher, P., Greenberg, J., Gunawardena, R., Gunawardena, R., Kwon, C.S., Tietje, O., Wong, C. (2006) Prediction of breast cancer using volatile biomarkers in the breath. *Breast cancer research and treatment*, 99 (1), 19-21.

Phillips, M., Cataneo, R. N., Saunders, C., Hope, P., Schmitt, P., Wai, J. (2010) Volatile biomarkers in the breath of women with breast cancer. *Journal of breath research*, 4 (2), 026003.

Pickel, D., Manucy, G. P., Walker, D. B., Hall, S. B. and Walker, J. C. (2004) Evidence for canine olfactory detection of melanoma. *Applied Animal Behaviour Science*. 89 (1-2), pp. 107-116.

Quignon, P., Giraud, M., Rimbault, M., Lavigne, P., Tacher, S., Morin, E., Retout, E., Valin, A-S., Lindblad-Toh, K., Nicolas, J., Galibert, F. (2005) The dog and rat olfactory receptor repertoires. *Genome biology*. 6 (10), R83.

Rooney, N. J., Morant, S. and Guest, C. (2013) Investigation into the value of trained glycaemia alert dogs to clients with type I diabetes. *PloS one*. 8 (8).

Rudnicka, J., Walczac, M., Kowalkowski, T., Jezierskib, T. and Buszewski, B. (2014) Determination of volatile organic compounds as potential markers of lung cancer by gas chromatography–mass spectrometry versus trained dogs. *Sensors and Actuators B: Chemical.* 202, pp. 615-621.

Smith, S., White, P., Redding, J., Ratcliffe, N. M., & Probert, C. S. (2010) Application of similarity coefficients to predict disease using volatile organic compounds. *IEEE Sensors Journal*, 10 (1), pp. 92-96.

Sonoda, H., Kohnoe, S., Yamazato, T., Satoh, Y., Morizono, G., Shikata, K., Morita, M., Watanabe, A., Morita, M., Kakeji, Y., Inoue, F. and Maehara, Y. (2011) Colorectal screening with odour material by canine scent detection. *Gut*. 60 (6), pp. 814-819.

Španěl, P., Smith, D., Holland, T. A., Singary, W. A., & Elder, J. B. (1999). Analysis of formaldehyde in the headspace of urine from bladder and prostate cancer patients using selected ion flow tube mass spectrometry. *Rapid communications in mass spectrometry*, *13*(14), 1354-1359.

Sreekumar, A., Poisson, L. M., Rajendrian, T. M., Khan, A. P., Cao, Q., Yu, J., Laxman, B., Mehra, R., Lonigro, R. J., Li, Y., Nyati, M. K., Ahsan, A., Kalyana-Sundaram, S., Han, B., Cao, X., Byun, J., Omenn, G. S., Ghosh, D., Pennathur, S., Alexander, D. C., Berger, A., Shuster, J. R., Wei, J. T., Varambally, S., Beecher, C. and Chinnaiyan, A. M. (2009) Metabolic profiles delineate potential role for sarcosine in prostate cancer progression. *Nature*. 457 (7231), pp. 910-914.

Taverna, G. Tidu, L. Grizzi, F. Torri, V. Mandressi, A. Sardella, P. La Torre, G. Cocciolone, G. Seveso, M. Giusti, G. Hurle, R. Santoro, A. Graziotti, P. (2015) Olfactory system of highly trained dogs detects prostate cancer in urine samples. *The Journal of Urology*. 193(4), pp. 1382-1387.

Tolonen, H., Kuulasmaa, K., Laatikainen, T., Wolf, H. (2002) Recommendation for indicators, international collaboration, protocol and manual of operations for chronic disease risk factor surveys, *European Health Risk Monitoring.* Part 3 section 4. http://www.thl.fi/publications/ehrm/product2/part_iii4.htm [accessed 16/02/2015].

Weber, C. M., Cauchi, M., Patel, M., Bessant, C., Turner, C., Britton, L. E., & Willis, C. M. (2011) Evaluation of a gas sensor array and pattern recognition for the identification of bladder cancer from urine headspace. *Analyst*, 136 (2), pp. 359-364.

Wickens, D. (2001) Remote air sampling for canine olfaction. *Security Technology, 2001 IEEE 35[th] International Carnahan Conference On.* pp. 100-102.

Williams, H. and Pembroke, A. (1989) Sniffer dogs in the melanoma clinic? *Lancet*. 1 (734).

Willis, C. M., Britton, L. E., Harris, R., Wallace, J., and Guest, C. M. (2011) Volatile organic compounds as biomarkers of bladder cancer: Sensitivity and specificity using trained sniffer dogs. *Cancer Biomarkers.* 8 (3), pp. 145-153.

Willis, C. M., Church, S. M., Guest, C. M., Cook, W. A., McCarthy, N., Bransbury, A. J., Church, M. R. T. and Church, J. C. T. (2004) Olfactory detection of human bladder cancer by dogs: proof of principle study. *British Medical Journal.* 329 (7468), pp. 712-715.